\documentclass[manuscript]{aastex}

\shorttitle{Comparison of Coronal Extrapolation Methods}
\shortauthors{Arden et al.}

\begin{document}

\title{Comparison of Coronal Extrapolation Methods for Cycle 24 Using HMI Data}

\author{William M. Arden}
\affil{University of Southern Queensland\\Toowoomba, Queensland, Australia}
%\email{U1045759@umail.usq.edu.au}

\author{Aimee A. Norton}
\affil{Hansen Experimental Physics Laboratory\\Stanford University, Stanford, CA, 94305 USA}
%\email{aanorton@stanford.edu}

\author{Xudong Sun}
\affil{Hansen Experimental Physics Laboratory\\Stanford University, Stanford, CA, 94305 USA}
%\email{xudongs@stanford.edu}

\and

\author{Xuepu Zhao}
\affil{Hansen Experimental Physics Laboratory\\Stanford University, Stanford, CA, 94305 USA}

\begin{abstract}
Two extrapolation models of the solar coronal magnetic field are compared using magnetogram data from the SDO/HMI instrument. The two models, a horizontal current--current sheet--source surface (HCCSSS) model and a potential field--source surface (PFSS) model differ in their treatment of coronal currents. Each model has its own critical variable, respectively the radius of a cusp surface and a source surface, and it is found that adjusting these heights over the period studied allows better fit between the models and the solar open flux at 1 AU as calculated from the Interplanetary Magnetic Field (IMF). The HCCSSS model provides the better fit for the overall period from 2010 November to 2015 May as well as for two subsets of the period --- the minimum/rising part of the solar cycle, and the recently-identified peak in the IMF from mid-2014 to mid-2015 just after solar maximum. It is found that a HCCSSS cusp surface height of 1.7 $R_\odot$ provides the best fit to the IMF for the overall period, while 1.7 \& 1.9 $R_\odot$ give the best fits for the two subsets. The corresponding values for the PFSS source surface height are 2.1, 2.2 and 2.0 $R_\odot$ respectively. This means that the HCCSSS cusp surface rises as the solar cycle progresses while the PFSS source surface falls. 
\end{abstract}


\section{INTRODUCTION}

Modeling solar coronal open flux is one of the tools available to solar physics in the search for understanding of the behavior of the corona. This is important, in turn, because of the corona's influence on space weather and the magnetosphere of the Earth. 

One class of these models involves extrapolation of coronal magnetic field and the interplanetary magnetic field (IMF) from the photospheric magnetogram. Early models achieved good results with simple, current-free models \citep{Schatten1969}; later versions include the effects of coronal currents \citep{Hoeksema1983}. At present, the Helioseismic and Magnetic Imager (HMI) of the Solar Dynamics Observatory (SDO) provides magnetograms at a high cadence, but photospheric magnetic data has been available for many years, beginning with the work of \citet{Hale1908} and continuing through the full-disk magnetograms of \citet{Babcock1963} in the 1950s \& 60s to current instrumentation such as HMI. Coronal models made use of this data as early as the 1960s \citep{Schatten1969}, and an improved form of these models (the Wang-Sheeley-Arge, or WSA model \citep{Arge2000}) is currently used by the NOAA Space Weather Prediction Center (SWPC) in forecasting the magnetospheric effects of solar activity.

Since these early developments, increases in computing power have enabled more sophisticated modeling of the corona through magnetohydrodynamic (MHD) approaches. Extrapolation models give comparable results \citep{Riley2006} and can be quickly and easily implemented with small-scale computing resources. Extrapolation models lend themselves to computing the long-term, relatively smooth quasi-static magnetic field at the corona. 

This paper compares two such extrapolation models, a potential field--source surface (PFSS) model developed at the Lockheed Martin Solar and Astrophysics Laboratory (LMSAL) \citep{Schrijver2003} and a horizontal current--current sheet--source surface (HCCSSS) model developed at Stanford University \citep{Zhao1994, Zhao1995}. Solar open magnetic flux calculated by the models is compared to open flux derived from the Interplanetary Magnetic Field (IMF) using a technique proposed by \citet{Lockwood2002}. It builds on earlier work by the authors \citep{Arden2014, Arden2015}. 

In this paper we examine HMI data for the period from 2010 November 26 to 2015 May 21, which encompasses the rising phase of solar cycle 24 and also the dramatic increase in solar mean field and open flux observed by \citet{Sheeley2015} starting in mid-2014. As those authors point out, this type of rise has characterized the declining phase of at least the last three solar cycles. This is particularly interesting since it has been shown that open flux during the declining phase is a reliable precursor of the activity in the next solar cycle \citep{Feynman1982}.

The paper begins with an outline of the data and analytical methods used. It continues with descriptions of the two models and the method of calculating open flux at 1 AU. The performance of the two models is compared and the results are discussed in the final sections of the paper.

\section{DATA \& METHODOLOGY}Coronal extrapolation models such as these begin with measured magnetic field data from photospheric observations. Field lines originating at the photosphere are mathematically extrapolated upwards to the corona and beyond, to an imaginary Òsource surfaceÓ outside of which all field lines are forced to be open and radial. Both models used in this paper follow this general approach, with differences described in detail in sections \ref{PFSS} and \ref{HCCSSS}. Table \ref{tbl-1} presents an abbreviated comparison of the two models.

A test of the accuracy of these models is comparison to the open flux at Earth, 1 AU from the Sun. The interplanetary magnetic field at Earth provides an \emph{in situ} measurement of open flux;  the IMF is measured by spacecraft such as ACE and made publicly available in NASA's OMNI 2 database.

\subsection{Magnetograms --- the Photospheric Field}\label{Magnetograms}The SDO/HMI instrument produces full-disk, line-of-sight magnetic images with a cadence of 45 seconds from the front camera and 12 minutes from the side camera. The 12-minute images are used from this study (HMI.M-720s data series, \url{http://jsoc.stanford.edu/jsocwiki/hmi.M-720s-info}). Even a full-disk image, however, only covers the Earth-facing half of the photosphere; accurate calculation of total open flux requires information about the entire photosphere. Therefore, a set of sequential images over a solar revolution is typically combined to form a synoptic map of the entire Sun. While these images do not represent the Sun at any one time (the earliest data incorporated into the synoptic map is approximately 27 days older than the latest data), they are sufficient as input for the quasi-static models discussed here.

In addition to the time-dependence of the measurements, the difficulty of measuring the magnetic field at large line-of-sight projection angles is well known \citep{Petrie2012}. These large angles occur at high solar latitudes, which makes them critical for accurate modeling; the unipolar magnetic regions at high latitudes comprise the polar caps where much of the high-speed solar wind originates. \citet{Sun2011} discuss the difficulties, and some possible solutions, to this problem.

Both of the models tested here are fundamentally based on spherical harmonic integration and thus require some estimate of the surface magnetic field over the entire photosphere. The two models use different methods to arrive at the magnetic fields at high latitudes (which are difficult to measure accurately). The LMSAL/SSW PFSS model uses a flux-dispersal model to estimate the polar magnetic field \citep{Schrijver2003}. The HCCSSS model incorporates polar field observations in fall \& spring (when the solar tilt angle is favorable) and estimates the values at other times through spatial-temporal interpolation \citep{Sun2011}.

This work uses synoptic maps from LMSAL as input to the PFSS model. These are publicly available at \url{http://www.lmsal.com/solarsoft/archive/ssw/pfss\_links\_v2/}. Synoptic maps used as input for the HCCSSS model were obtained from Stanford University \citep{Sun2011}.

\subsection{Potential Field Source Surface (PFSS) model}\label{PFSS}Using the solar magnetic field measured at the photosphere as the lower boundary condition for a coronal model is well established. Beginning with early work by \citet{Altschuler1969} and \citet{Schatten1969} and continuing through work by \citet{Schrijver2003} and many others, these models have reached a level of sophistication that enables them to be used for near-real time space weather prediction by the NOAA SWPC  \citep{Arge2000} (see, for example, \url{http://www.swpc.noaa.gov/products/wsa-enlil-solar-wind-prediction}). \citet{Mackay2012} provide an overview of many of these models.

\begin{figure}
	\centering
	\includegraphics[width=20pc]{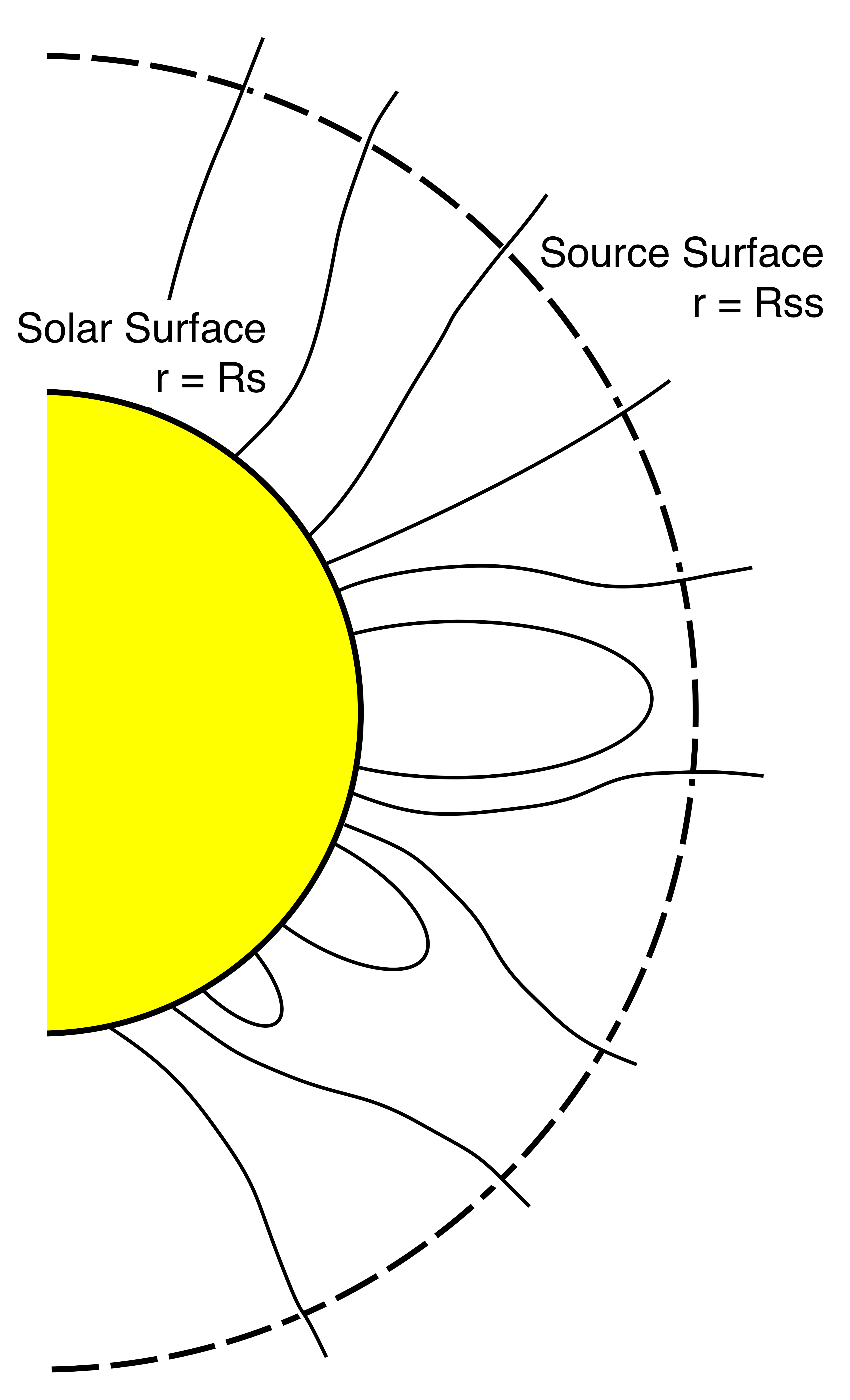}
	\caption{\label{PFSS_cartoon} Diagrammatic representation of PFSS model. The PFSS model assumes a ``source surface" at which all magnetic field lines are open and radial. Moving the PFSS source surface to a lower height increases the open flux because it forces more higher-order structures to become open.}
\end{figure}

The corona is known to carry electric currents, but these currents are generally not significant on the global scale except near the heliospheric current sheet. The lower boundary condition for the model is the photospheric field derived from magnetograms (with radius equal to that of the Sun, $R_\odot$). This field depends on radial distance as well as solar latitude and longitude ($R_\odot = B(r,\theta, \phi)$). An imaginary sphere called the source surface at radius $R_{ss}$, where the magnetic field is assumed to be purely radial ($R_{ss} = B(r)$), provides the upper boundary condition. As its name implies, the PFSS technique does not consider coronal currents below the source surface. Figure \ref{PFSS_cartoon} shows a schematic of the PFSS model.

Under these constraints,  $\mathbf{\nabla}\times\mathbf{B} = 0$ and $\mathbf{B} = -\nabla\Psi$, where $\Psi$ is a scalar potential (see \citet{Mackay2012}). $\Psi$ satisfies Laplace's equation:

\begin{equation}
\nabla^2\Psi = 0
\end{equation}
with boundary conditions
\begin{equation}
\left.\frac{\partial\Psi}{\partial r}\right|_{r=R_s} = -B_r(R_s, \theta, \phi) 
\end{equation}
and
\begin{equation} \left.\frac{\partial\Psi}{\partial \theta}\right|_{r=R_{ss}} = \left.\frac{\partial\Psi}{\partial \phi}\right|_{r=R_{ss}} = 0.  
\end{equation}
This equation is solved using spherical harmonic methods to give the components of $B(r, \theta, \phi)$ at any point over the range $R_s \le r \le R_{ss}$.

The PFSS model used in the present study was developed at LMSAL. It uses the SolarSoft (IDL) software package and assimilates surface-flux estimates along with MDI and HMI magnetogram data. This model is currently in its second revision (\url{http://www.lmsal.com/forecast/}).

The output of a PFSS model includes values of the radial B field ($B_r$, in nT) over a grid at the source surface. Total open flux is obtained by integrating the absolute values of the radial field over the sphere and then dividing by two (simply integrating the B field would result in zero net flux). In actual computation, this integration is half the sum of the absolute values weighted by the area of the respective grid elements.

Synoptic magnetograms of the photosphere at a cadence of one per day are used as input to the PFSS model. The output is a daily file of B field as described above.

\subsection{Varying the PFSS source surface for better fit}
The source surface height $R_{ss}$ is commonly taken to be 2.5 $R_\odot$ \citep{Hoeksema1983}. In fact, though, this is a free parameter in the PFSS model. Varying the height of the source surface over the course of the solar cycle has been explored by \citet{Lee2011} and \citet{Arden2014}. Adjusting $R_{ss}$ is one way to better match the open flux measured at 1 AU. After examination of source surface heights ranging in value from 1.5 to 2.5 $R_\odot$, the values chosen for this study are 2.0 and 2.25 $R_\odot$ for different phases of the solar cycle.

The connection between source surface height and open flux is this: Smaller spatial-scale magnetic loops that close lower in the atmosphere are represented by the higher spherical harmonic orders. Loops that close beneath the source surface (and therefore do not penetrate it) do not contribute to the open flux above the source surface. As the source surface height is raised, therefore, fewer of the higher order loops penetrate the source surface and open flux decreases. As source surface height is lowered, more of these higher spatial-order loops cross the source surface and open flux increases.

In \citet{Arden2014}, it was shown that moving the source surface to higher values in the PFSS model gives a better fit as the solar cycle passed from maximum to minimum. The results of the current study support that conclusion (see \ref{Discussion}). 

\subsection{Horizontal Current -- Current Sheet -- Source Surface (HCCSSS) model}\label{HCCSSS}
The HCCSSS model takes a different approach to modeling the corona. This model begins with the assumption of a corona in magnetohydrostatic equilibrium with horizontal electric currents instead of a potential field, and adds a cusp surface to model the effect of streamer current sheets \citep{Zhao1994, Zhao1995}. This model, unlike PFSS, is not force-free due to the inclusion of pressure and gravity. Finally, the HCCSSS model adds a source surface to include volume currents beyond the cusp surface. The resulting model gets its name from the inclusion of horizontal currents, current sheet, and a source surface. Note that in \citet{Zhao1995}, this model is called CSSS; the name HCCSSS is more descriptive and will be used here. 

\begin{figure}
	\centering
	\includegraphics[width=20pc]{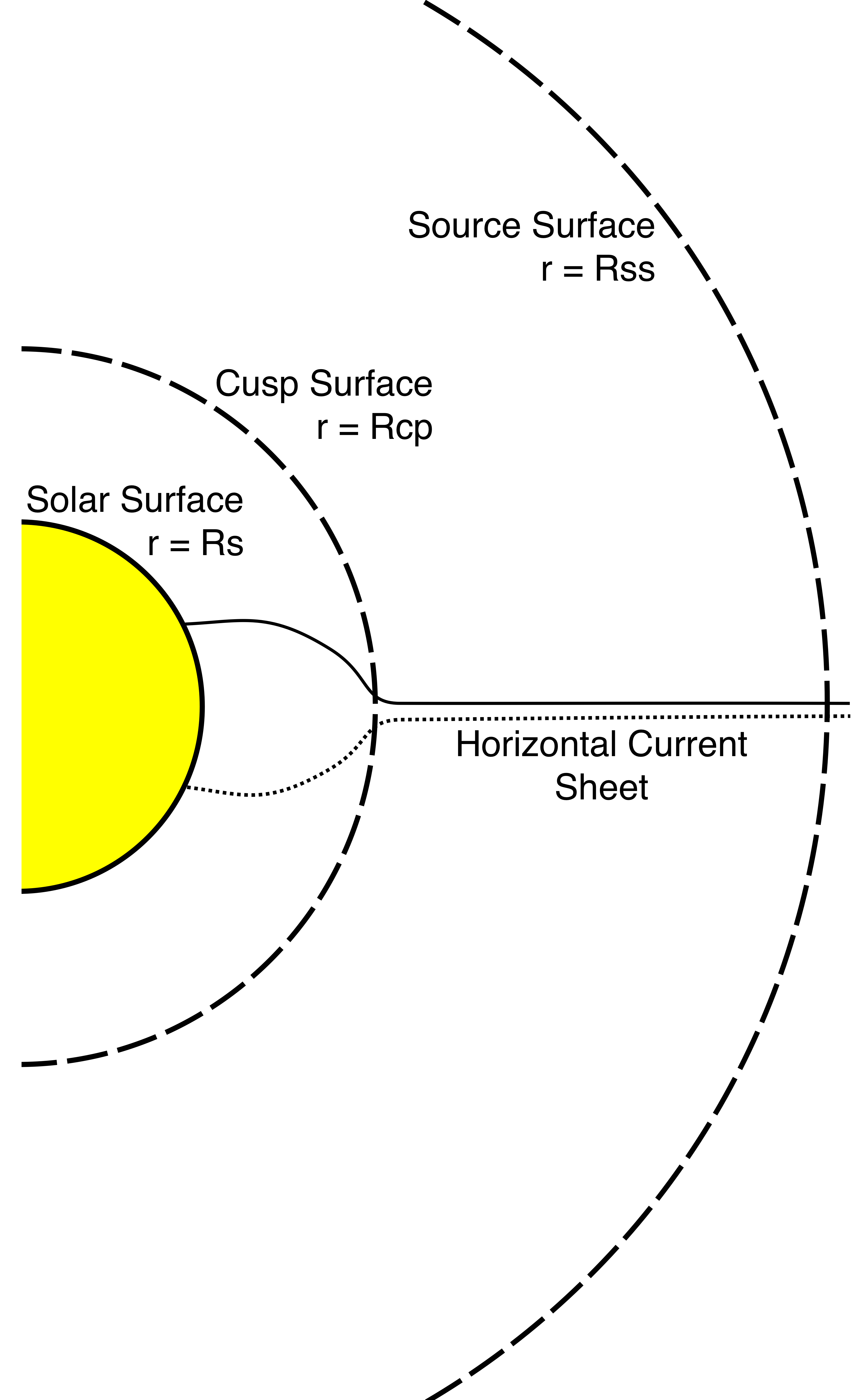}
	\caption{\label{HCCSSS_cartoon} Diagrammatic representation HCCSSS model. The HCCSSS model includes a ``cusp surface," explained in the text. Moving this cusp surface in or out affects the open flux calculated by this model.}
\end{figure}

As described by Zhao \& Hoeksema, helmet-streamer structures such as those observed in solar eclipses indicate that coronal currents alter the magnetic topology. A streamer interface starts near the cusp-shaped neutral point over a closed region. Coronal currents are assumed to have two components: horizontal volume currents in the corona and sheet currents flowing within streamer interfaces. All helmet-streamer components are assumed to have identical heights ($r=R_{cp}$, the ``cusp height"). The corona is then divided into two regions separated by a spherical surface located at $R_{cp}$, called the ``cusp surface." Magnetic field lines are assumed to be closed below the cusp surface and open (but not necessarily radial) above it. The outer region is bounded by a ``source surface" corresponding to the similarly named surface of the PFSS model, at which the magnetic field lines are both open and radial. Figure \ref{HCCSSS_cartoon} depicts the HCCSSS model. See Table \ref{tbl-1} for a comparison of the PFSS and HCCSSS models.

The HCCSSS model \citep{Zhao1994} begins with the equation for magnetohydrostatic force balance in $1/r^2$ gravity \citep{Bogdan1986}:

\begin{equation}
\frac{1}{4\pi}(\mathbf{\nabla} \mathbf{\times} \mathbf{B}) \mathbf{\times} \mathbf{B} - \mathbf{\nabla} p - \rho \frac{GM}{r^2} \mathbf{\hat{r}} = 0
\end{equation}

where $\mathbf{B}$ is the magnetic field, p the plasma pressure and $\rho$ the plasma density. \citet{Bogdan1986} found that this equation has a set of solutions that depend on a function $\Phi(r,\theta, \phi)$:

\begin{equation}
\mathbf{B} = -\eta (r) \frac{\partial\Phi}{\partial r}\hat{r}-\frac{1}{r}\frac{\partial\Phi}{\partial\theta}\hat{\theta}-\frac{1}{r\sin\theta}\frac{\partial\Phi}{\partial\phi}\hat{\phi}
\end{equation}

\begin{equation}
p = p_0 (r) + \frac{1}{8\pi}\eta (r)[1-\eta (r)]\left( \frac{\partial \Phi}{\partial r}\right)^2
\end{equation}

\begin{equation}
\rho=\rho_0(r) + \frac{1}{GM}\left\{\frac{\eta(r)-1}{8\pi} \frac{\partial}{\partial r}{\left(\frac{\partial\Phi}{\partial\theta}\right)}^2+\frac{1}{8\pi} \frac{\eta (r)-1}{\sin^2\theta} \frac{\partial}{\partial r}{\left(\frac{\partial\Phi}{\partial\phi}\right)}^2+\frac{r^2}{8\pi} \frac{\partial}{\partial r}\left[\eta(r)[\eta (r)-1]{\left(\frac{\partial\Phi}{\partial r}\right)}^2\right]\right\}
\end{equation}

where

\begin{equation}
\eta(r) = {\left(1+\frac{a}{r}\right)}^2
\end{equation}

Solutions for each of the two regions are formulated, with the constraint that all three components of the magnetic field be continuous across the cusp surface. Further details can be found in \citet{Zhao1994}. In actual computation, only \emph{a} and $R_{cp}$, along with the observed photospheric magnetic field, are required to calculate the magnetic field above the photosphere. In this paper, the value of \emph{a} is chosen to be 0.2, as described in \citet{Zhao1995}. It is below this height that the horizontal currents primarily flow. The model is relatively insensitive to the choice of $R_{ss}$, the source surface height (which corresponds to the base of the heliosphere); values of 10-15 are tested in this research with little difference in the final results for open flux. In accordance with \citet{Zhao2010}, a value of 15 $R_\odot$ was chosen. Cusp surface heights of 1.5, 1.7, 2.0 and 2.2 $R_\odot$ were explored.

In its present form, the HCCSSS model takes as input synoptic maps of the photosphere at a cadence of one Carrington Rotation (CR). The output is thus an average of open flux over that CR. 

In a similar technique to that employed in the PFSS model, the open flux calculated by the HCCSSS model can be adjusted by moving the cusp surface higher (which decreases open flux) or lower (which increases open flux).

\subsection{Open Flux at 1 AU --- IMF at Earth} 
Thanks to the work of \citet{Lockwood2013} and others, there is a direct way to arrive at an estimate of open magnetic flux at 1 AU. It is based on values of $B_r$, which are available from the OMNI 2 database (\url{http://omniweb.gsfc.nasa.gov/ow.html}). This database was created in 2003 and contains \emph{in situ} solar wind magnetic field and plasma data from a number of near-Earth spacecraft at a one-hour cadence. It includes IMF data from 1963 November through the present; all three components of the field ($B_x$, $B_y$ and $B_z$) are given, in units of nT. We take $B_x$, the magnetic field component along the Sun-Earth line, to be equal to the radial field, and use data from 2010 November through 2015 May. Daily averages, which are also found in the OMNI 2 database, are used in our calculations and then averaged by Carrington Rotation to correspond to the output of the HCCSSS model.

The radial component of the heliospheric magnetic field has been shown to be independent of heliospheric latitude; this was deduced from measurements by the  \emph{Ulysses} spacecraft \citep{Smith1995, Smith2008, Smith2011}. Based on this assumption, the total unsigned flux passing through a sphere with a radius of 1 AU ($R_1$) can be given simply by 
\begin{equation}
\label{lockwood}
F = 4 \pi R_{1}^2 |B_{r}|/2
\end{equation}
as shown by \citet{Lockwood2002}. Note that the factor of two is required, for the following reason. If the total open flux over a sphere at 1 AU is calculated from the IMF measured at Earth, all the flux over the whole sphere will be presumed to be of that sign. Taking the absolute value of $B_{r}$ removes the sign of the polarity, but effectively makes all the flux positive. The net open flux must be zero, so division by two is required since both the positive and negative flux would otherwise be counted as positive --- resulting in a value that is twice the actual one. In the paper, we chose to call this flux ``unsigned" since it represents the flux of both polarities.

\subsection{Comparing IMF to Modeled Open Flux}
We compare the calculated IMF at 1 AU with the results of the PFSS and HCCSSS models over the time period from 2010 November 26 through 2015 May 21, which corresponds to Carrington Rotations (CR) 2104--2163. As described earlier, the HCCSSS model is currently based on synoptic maps of one CR. The PFSS and IMF data are therefore binned by CR by averaging daily values over corresponding periods. Goodness of fit is determined by calculating the RMS value of the difference between the open flux derived from the IMF and each model over all the CRs in the period.

In addition to comparing the IMF to the models over the entire period, comparisons are made for two epochs within the period. From the beginning of 2010 through mid-2014, the Sun appeared to be experiencing a typical, if low activity, cycle. In mid-2014, however, there was a dramatic rise in the magnitude of the Sun's large-scale magnetic field due to a significant increase in the Sun's dipole moment \citep{Sheeley2015}. While Sheeley \& Wang point out that this pattern has occured after solar maximum in each of the three previous solar cycles, it is not as widely known as other aspects of the solar cycle; it does not appear in common measures of solar activity such as sunspot count, for example. During this period, the strength of the IMF radial component doubled; that increase is readily apparent in our data, and it seemed appropriate to break our tests of the models into two parts --- one for the relatively calm period from late 2010 through mid-2014, and one for the rise (and subsequent fall) of the IMF from mid-2014 to mid-2015.

\section{RESULTS}

Figure \ref{double_plot} shows unsigned open flux computed from the IMF compared to the outputs of the HCCSSS and PFSS models, each at two different cusp surface or source surface heights. In these plots, the average value of the IMF-based or modeled open flux is plotted using one point per CR for visual comparison. At a cursory glance, both models track at least the gross features of the IMF when averaged at one CR, including the 2014-2015 peak in the IMF. The HCCSSS model with a cusp surface height of $1.7 R_\odot$ appears to follow the IMF better from the beginning of the period up to the middle of 2014 (when the IMF rose dramatically) but then overestimates the open flux during the IMF peak. The PFSS model gives varying results; a source surface height of 2.2 $R_\odot$ gives a closer visual fit over the earlier part of the period, but 2.0 $R_\odot$ appears closer in the later part (the peak of cycle 24).

For a more quantitative measure, we calculate the difference between the modeled open flux and the IMF open flux, one CR at a time, and then compute the RMS value of the collective differences. These values are shown in Table \ref{rms_diff} for three different subsets of the data --- the entire period from 2010 November 26 to 2015 May 21 (CR 2104--2163), the quasi-steady state period from 2010 November 26 to 2014 July 24 (CR 2104--2152) and the period of the IMF's surge from 2014 July 25 to 2015 May 21 (CR 2153--2163). 

The results in Table \ref{rms_diff} can be summarized as follows:
\begin{itemize}
\item Overall, a cusp surface height of $1.7 R_\odot$ in the HCCSSS model gives the best fit of either model for the entire period under study. Setting the source surface height to $2.1 R_\odot$ in the PFSS model gives the best fit for that model, but the fit is not as good as that achieved with the HCCSSS model.
\item For the relatively uneventful first epoch, CR 2104--2152, the HCCSSS model with a cusp surface height of $1.7 R_\odot$ again gives a better fit than any of the three PFSS source surface heights. The best PFSS source surface height is $2.2 R_\odot$.
\item During the peak in the IMF, CR 2153--2163, the HCCSSS model with a $1.7 R_\odot$ cusp surface height overestimates the IMF open flux; raising the cusp surface height to $1.9 R_\odot$ lowers the model's open flux and provides a better fit. On the other hand, the PFSS model underestimates the IMF open flux; lowering the source surface height from $2.2 R_\odot$ to $2.0 R_\odot$ raises the model's open flux to more closely match the IMF open flux. 
\end{itemize}

\begin{figure}
\rotate
\includegraphics[width=37pc]{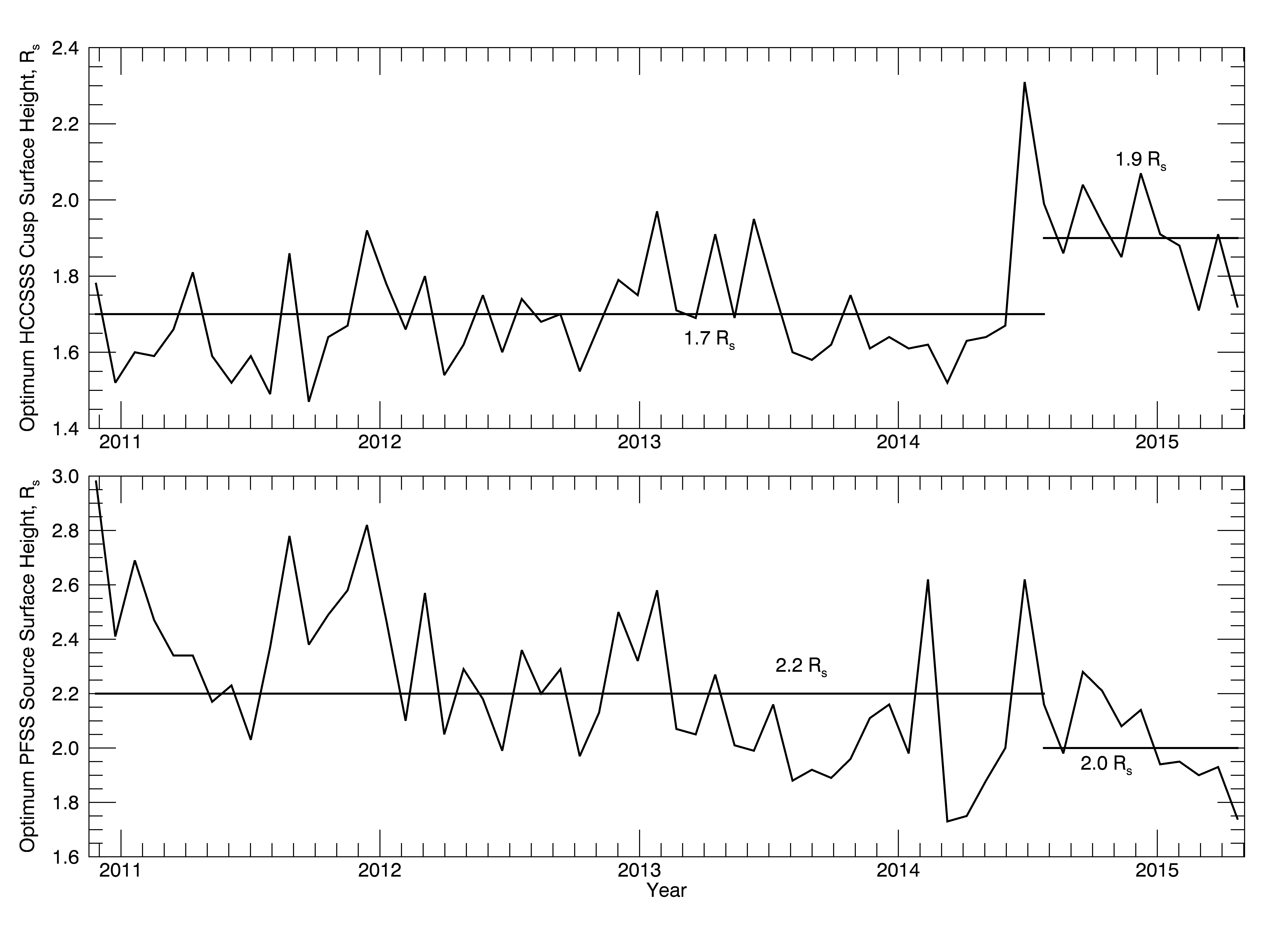}
\caption{\label{optimum_plot} Interpolated optimum HCCSSS cusp surface heights (top plot) and PFSS source surface heights (bottom plot), in units of solar radii. These values give the best match to the IMF open flux for each CR. Also shown are reference lines corresponding to the overall values chosen for each epoch on the basis of minimum RMS difference between IMF and coronal model.}
\end{figure}

Figure \ref{optimum_plot} illustrates the variation in optimal cusp surface and source surface heights. Here, interpolated values of the cusp surface and source surface heights that give the best fit to the IMF open flux are shown for each CR. Reference values of 1.7 $R_\odot$ and 1.9 $R_\odot$ are shown for the HCCSSS model as well as values of 2.0 $R_\odot$ and 2.2 $R_\odot$ for the PFSS model, in their corresponding epochs. OMNI data from 2014 July shows a drop in IMF open flux just before the striking increase which resulted in the to-date solar cycle 24 maximum IMF value of $5.2x10^{22}$ Mx in 2014 November. The first and second epochs are divided so that 2014 July is included in the first epoch (CR 2104-2152) because the IMF did not reach values higher than average until 2014 August. For the HCCSSS model (upper plot, figure \ref{optimum_plot}), the overall mean cusp surface height of the interpolated values is 1.7 $R_\odot$. For the first epoch, the mean height is also 1.7 $R_\odot$ while for the second epoch it is 1.9 $R_\odot$. The corresponding mean source surface heights for the PFSS model (lower plot, figure \ref{optimum_plot}) are 2.1, 2.2 and 2.0 respectively. All of these values are represented in table \ref{rms_diff}.

\section{DISCUSSION \& CONCLUSIONS}\label{Discussion}

The ideal coronal extrapolation model would allow initial specification of parameters such as source surface height or cusp surface height, and would then reproduce coronal behavior precisely as the photospheric inputs change. Attempts to develop such a model lead to open questions as well as open flux --- What are the optimum heights of the cusp surface and source surface, and when should they be changed? Why does a model work well under some circumstances but need modification in others? The recent work of \citet{Sheeley2015} describing an unexpected rise in the IMF in late 2014 provides us the opportunity to study these models in both quasi-steady state and widely varying situations. We find that both of the models require modification of a primary variable in order to track the IMF accurately in both the rising phase (2010 -- 2014) and the post-maximum IMF peak; cusp surface height in the case of the HCCSSS model, and source surface height for the PFSS model. 

Over the course of the period studied, the mean value of the CR-averaged IMF was $2.9x10^{22}$ Mx. Table \ref{rms_diff} includes the percent error (100\% x RMS difference / mean IMF) for the values of cusp surface and source surface. It is clear that the HCCSSS model gives at least a slightly better, and in some cases significantly better, fit than the PFSS model in all cases; the smallest errors for the entire period, the rising phase of cycle 24, and the IMF peak in the declining phase all result from the use of the HCCSSS model. Both models, however, are capable of errors of 20\% or less with appropriate tuning of source surface / cusp surface heights.

The results of this study can be compared to the conclusions of an earlier article by the authors \citep{Arden2014}. In that work, it was found that a higher PFSS source surface during solar minimum resulted in a better fit to open flux at 1 AU as calculated from the IMF. The source surface was lowered as maximum approached, which improved the fit. The NOAA SWPC estimates that the current solar cycle began in early 2009 and reached its peak in the first half of 2014 (\url{http://www.swpc.noaa.gov/products/solar-cycle-progression}). The current study begins in 2010, soon after minimum, and we find that a higher source surface value does, indeed, give a better fit for the first part of the study (2010 November -- 2014 July) --- the minimum and rising phases of cycle 24. Lowering the source surface improves the fit during the period from 2014 July to 2015 May --- the period of solar maximum and the peak in the IMF, and the end of this study.

We find that the opposite is true for the HCCSSS model, whose critical parameter is the height of the cusp surface and which introduces a more sophisticated treatment of coronal currents. It is similar to the PFSS model in that raising the cusp surface lowers calculated open flux. Since the model overestimates the open flux during the IMF peak, therefore, raising the cusp surface as maximum approached yields a better fit to the IMF open flux.

\begin{figure}
\rotate
\includegraphics[width=37pc]{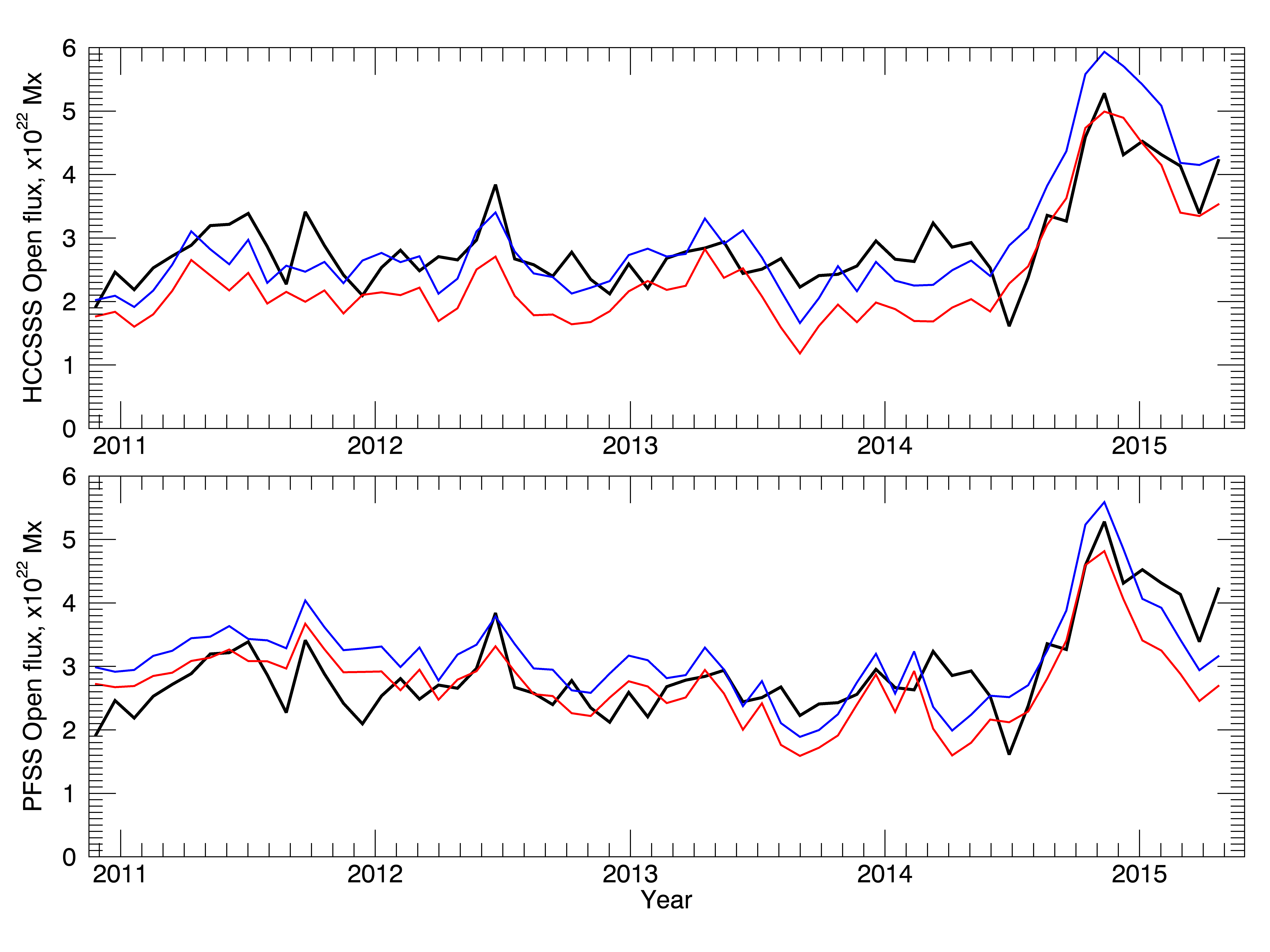}
\caption{\label{double_plot} IMF, HCCSSS and PFSS unsigned open flux. In both plots, the heliospheric open flux calculated from the IMF OMNI 2 database is plotted by the black line. Top: HCCSSS open flux is plotted against IMF open flux for HCCSSS cusp surface heights of $1.7 R_\odot$ (upper line, blue) and $1.9 R_\odot$ (lower line, red). Bottom: PFSS open flux at source surfaces of 2.0 $R_\odot$ (upper line, blue) and 2.2 $R_\odot$ (lower line, red) are plotted against IMF open flux. IMF data from the OMNI 2 database.}
\end{figure}

The choice of epochs notwithstanding, examination of Figure \ref{double_plot} reveals that the starting HCCSSS cusp surface of $1.7 R_\odot$ provides a better fit for a longer time (2010 to mid-2014) than the starting PFSS source surface height of $2.25 R_\odot$, which begins to deviate significantly from the IMF open flux in mid-2013 --- as maximum approaches. In other words, the time at which the source surface needs to be lowered (mid-2013) is distinctly different from the time that the cusp surface needs to be raised (mid-2014).

This paper examines the effect of changing HCCSSS cusp surface and PFSS source surface heights. With regard to the HCCSSS model in particular, it is expected that all three free parameters of the model (the variable \emph{a}, cusp surface height and source surface height) probably vary over the solar cycle, and this variation could significantly affect the calculated results. It is appropriate to continue this study to find the optimum values of all three parameters. Also, while there is no physical surface against which the PFSS source surface height can be tested, validation of the average HCCSSS cusp surface height reported here by comparison to the cusp height of streamers as observed in coronagraph data could be a profitable avenue for future exploration.

We have focused here on the behavior of two models and demonstrated that critical parameters in each model must be adjusted to fit the IMF at Earth. We have not addressed the reasons why these adjustments are necessary. We believe that the answer lies largely in the phenomena that affect the solar magnetic field on its way from the Sun to 1 AU, includng coronal mass ejections (CME) \citep{Owens2006, Owens2011, Schwadron2010}. These phenomena are not modeled by either technique --- nor, for that matter, by any method based on extrapolation of the photospheric magnetic field. While the authors believe that effects of most of these transient effects are averaged out over time periods on the order of one CR, this remains very much an area for further research and discussion. 

Allowing the parameters of a model to change over the course of the solar cycle enables us to more closely model the IMF open flux in retrospect, but at the cost of finding a single value that could be used predictively in fields such as space weather forecasting. The search continues for a model which captures the most significant aspects of the corona's complex and dynamic behavior without needing to adjust surface heights or other variables. 

\acknowledgments

The authors wish to acknowledge J. Todd Hoeksema of Stanford University and Marc DeRosa of LMSAL for their assistance in the research that led to this paper. The OMNI data were obtained from the GSFC/SPDF OMNIWeb interface at \url{http://omniweb.gsfc.nasa.gov}.

%\clearpage

%\clearpage

%\clearpage

%\clearpage

\begin{deluxetable}{p{4cm}p{5.5cm}p{5.5cm}}
\tabletypesize{\scriptsize}
\tablecaption{Comparison of Extrapolation-Based Coronal Magnetic Field Models.\label{tbl-1}}
\tablehead{
\colhead{Description} & \colhead{PFSS} & \colhead{HCCSSS}
}
\startdata
\setlength{\baselineskip}{-4pt}
Abbreviation & Potential Field Source Surface & \raggedright Horizontal Current -- Current Sheet -- Source Surface\raggedright \tabularnewline
\tableline
Lower Boundary Condition & \multicolumn{2}{c}{Observed Photospheric Magnetic Field}\\
\tableline
Upper Boundary Condition & Source surface ($\approx 2.5 R_\odot$) only & \raggedright Cusp surface ($1.7-2.0 R_\odot$) \& \\Source surface ($\approx 10-15 R_\odot$)\tabularnewline
\tableline
Assumptions & 
\setlength{\baselineskip}{-3pt}
\begin{itemize}
\vspace{-18pt}\setlength{\itemindent}{-20pt}
\item Force-free
\item No coronal currents below source surface
\item Field lines open \& radial at source surface
\end{itemize} & 
\setlength{\baselineskip}{-3pt}
\begin{itemize}
\vspace{-18pt}\setlength{\itemindent}{-20pt}
\item Magnetohydrostatic, not force-free
\item Horizontal volume currents in corona, sheet currents within streamer interfaces
\item Field lines open at cusp surface, radial at source surface
\end{itemize}\\
\tableline
\setlength{\baselineskip}{-3pt}
Theoretical foundation & Purely potential field; \begin{math} \mathbf{B}=-\mathbf{\nabla}\mathbf{\Psi}\end{math} where \begin{math} \mathbf{\nabla}^2 \mathbf{\Psi}=0 \end{math} (Green's function solution to Maxwell's equations) & \vspace{-18pt}\flushleft Magnetohydrostatic equilibrium; \begin{math}\frac{1}{4\pi}(\mathbf{\nabla x B}) \mathbf{x B} - \mathbf{\nabla}p - \rho\frac{GM}{r^2}\mathbf{\hat{r}}=0  \end{math}\tabularnewline
\tableline
\setlength{\baselineskip}{-3pt}
Source & \citet{Schatten1969} & \citet{Low1985}; \citet{Bogdan1986}\\
\tableline
\setlength{\baselineskip}{-3pt}
References & \citet{Schatten1969} & \citet{Zhao1995}\\
\enddata
\end{deluxetable}

\begin{deluxetable}{llccccc}
\tabletypesize{\tiny}
%%\tabletypesize{\scriptsize}
\tablecaption{\label{rms_diff} RMS value of (IMF $-$ coronal model) difference and (percent difference) between mean IMF and model, over CR 2104-2163.}
\tablehead{
	\colhead{} & 
		\colhead{} &
		\multicolumn{2}{c}{IMF $-$ HCCSSS} &
		\multicolumn{3}{c}{IMF $-$ PFSS}\\
	\colhead{} &
		\colhead{} &
		\multicolumn{2}{c}{(RMS x$10^{21}$ Mx)} & 
		\multicolumn{3}{c}{(RMS x$10^{21}$ Mx)}\\
	\colhead{} &
		\colhead{} &
		\multicolumn{2}{c}{Cusp surface height} &
		\multicolumn{3}{c}{Source surface height}\\
	\colhead{Period (CR)} &
		\colhead{Dates} &
		\colhead{1.7 $R_\odot$} & \colhead{1.9 $R_\odot$} & \colhead{2.0 $R_\odot$} & \colhead{2.1 $R_\odot$} & \colhead{2.2 $R_\odot$}
}
\startdata
	\tableline 2104--2163 & 2010 Nov 26 -- 2015 May 21 & \bfseries{5.31 (18\%)} & 7.00 (24\%) & 5.77 (20\%) & 5.4 (19\%) & 5.76 (20\%)\\
	\tableline 2104--2152 & 2010 Nov 26 -- 2014 Jul 24 & \bfseries{4.39 (15\%)} & 7.52 (26\%) & 5.79 (20\%) & 5.07 (18\%) & 4.97 (17\%)\\
	\tableline 2153--2163 & 2014 Jul 25 -- 2015 May 21 & 8.28 (29\%) & \bfseries{4.08 (14\%)} & 5.90 (20\%) & 6.88 (24\%) & 8.57 (30\%)\\
\enddata
\end{deluxetable}


\begin{thebibliography}{}

\bibitem[Altschuler \& Newkirk(1969)]{Altschuler1969}
Altschuler, M.D., \& Newkirk, Jr., G. 1969, SoPh, 9, 131

\bibitem[Arden et al.(2014)]{Arden2014}
Arden, W.M., Norton, A.A. \& Sun, X. 2013, JGRA, 119, 1476

\bibitem[Arden \& Norton(2015)]{Arden2015}
Arden, W.M. \& Norton, A.A. 2015, \textit{Triennial Earth-Sun Summit} (poster presentation)
  
\bibitem[Arge \& Pizzo(2000)]{Arge2000}
Arge, C.N. \& Pizzo, V.J. 2000, JGR, 105, 10465

\bibitem[Babcock(1963)]{Babcock1963}
Babcock, H.W. 1963, ARA\&A, 1,41

\bibitem[Bogdan \& Low(1986)]{Bogdan1986}
Bogdan, T.J. \& Low, B.C. 1986, \apj, 306,271
  
\bibitem[Feynman(1982)]{Feynman1982}
Feynman, J. 1982, JGR, 87, 6153

\bibitem[Hale(1908)]{Hale1908}
Hale, G. E. 1908, Mount Wilson Solar Observ. Contrib., 26
  
\bibitem[Hoeksema et al.(1983)]{Hoeksema1983}
Hoeksema, J.T., Wilcox, J.M. \& Scherrer, P.H. 1983, JGR, 88, 9910

\bibitem[Lee et al.(2011)]{Lee2011}
Lee, C.O., Luhmann, J.G., Hoeksema, J.T., Sun, X., Arge, C.N., \& dePater, I. 2011, SoPh, 269, 367

\bibitem[Lockwood(2002)]{Lockwood2002}
Lockwood, M.  2002 in From Solar Min to Max: Half a Solar
  Cycle with SOHO, ESA SP-508, SOHO 11 Symposium, Davos, Switzerland.
  
\bibitem[Lockwood(2013)]{Lockwood2013}
Lockwood, M. 2013, LRSP, 10, 4

\bibitem[Low(1985)]{Low1985}
Low, B.C. 1985, ApJ, 293, 31

\bibitem[Mackay \& Yeates(2012)]{Mackay2012}
Mackay, D.H., \& Yeates, A.R. 2012, LRSP, 9, 6

\bibitem[Owens \& Crooker(2006)]{Owens2006}
Owens, M.J. \& Crooker, N.U. 2006, JGR, 111, A10104

\bibitem[Owens et al.(2011)]{Owens2011}
Owens, M.J., Crooker, N.U. \& Lockwood, M. 2011, JGR, 116, A04111 
  
\bibitem[Petrie(2012)]{Petrie2012}
Petrie, G.J.D. 2012, SoPh, 281, 577

\bibitem[Riley(2006)]{Riley2006}
Riley, P., Linker, J.A., Mikic, Z., \& Lionello, R. 2006, ApJ, 653, 1510

\bibitem[Schatten et al.(1969)]{Schatten1969}
Schatten, K.H., Wilcox, J.M. \& Ness, N.F. 1969, SoPh, 9, 442

\bibitem[Schrijver \& DeRosa(2003)]{Schrijver2003}
Schrijver, C.J., \& DeRosa, M.L. 2003, SoPh, 212, 165

\bibitem[Schwadron et al.(2010)]{Schwadron2010}
Schwadron, N.A., Connick, D.E. \& Smith, C. 2010, ApJL, 722, L132

\bibitem[Sheeley \& Wang(2015)]{Sheeley2015}
Sheeley, Jr., N.R. \& Wang, Y.-M. 2015, ApJ, 809, 113
  
\bibitem[Smith \& Balogh(1995)]{Smith1995}
Smith, E.J., \& Balogh, A. 1995, GeoRL, 22, 3317
 
\bibitem[Smith(2008)]{Smith2008}
Smith, E.J. 2008, in The Heliosphere Through the Solar Activity Cycle, ed. A. Balogh, L.J. Lanzerotti \& S.~T. Seuss, pp. 79, Praxis, Chichester, U.K. 
 
\bibitem[Smith(2011)]{Smith2011}
Smith, E.J. 2011, JASTP, 73, 277.

\bibitem[Sun et al.(2011)]{Sun2011}
Sun, X., Liu, Y., Hoeksema, J.T., Hayashi, K., \& Zhao, X. 2011, SoPh, 270, 9
  
\bibitem[Zhao \& Hoeksema(1994)]{Zhao1994}
Zhao, X. P. \& Hoeksema, J.T. 1994, SoPh, 151, 91
  
\bibitem[Zhao \& Hoeksema(1995)]{Zhao1995}
Zhao, X. P. \& Hoeksema, J.T. 1995, JGR, 100, 19

\bibitem[Zhao \& Hoeksema(2010)]{Zhao2010}
Zhao, X. P. \& Hoeksema, J.T. 2010 SoPh, 266, 379

\end{thebibliography}
\end{document}